\documentclass[12pt,aps,showpacs,showkeys,preprintnumbers,amsmath,amssymb]{revtex4}

\usepackage{graphicx}

\def\pmb#1{\setbox0=\hbox{$#1$}%
  \kern-.025em\copy0\kern-\wd0
  \kern.05em\copy0\kern-\wd0
  \kern-.025em\raise.0433em\box0}

\def\beq{\begin{equation}}
\def\eeq{\end{equation}}

\begin{document}

\title{Observers in Spacetime and Nonlocality}

\author{B. Mashhoon}
  \affiliation{Department of Physics and Astronomy, University of Missouri, Columbia, Missouri 65211, USA}

\date{\today}

\begin{abstract}
Characteristics of observers in relativity theory are critically examined. For field measurements in Minkowski spacetime, the Bohr-Rosenfeld principle implies that the connection between actual (i.e., noninertial)  and inertial observers must be nonlocal. Nonlocal electrodynamics of non-uniformly rotating observers is discussed and the consequences of this theory for the phenomenon of spin-rotation coupling are briefly explored.  
\end{abstract}



\maketitle
\section{Introduction}

In the standard theory of relativity, Lorentz invariance---i.e., the  invariance of the laws of microphysics under the group of passive inhomogeneous Lorentz transformations---is extended to accelerated systems and gravitational fields in a pointwise manner~\cite{0}. Thus phenomena involving classical point particles and rays of radiation can be naturally described in terms of pointlike \emph{coincidences} in spacetime. On the other hand, electromagnetic fields cannot be measured instantaneously. It is argued here that these contrary elements can be brought together in a \emph{nonlocal} framework. 

An important consequence of the Lorentz invariance of Maxwell's equations is that the propagation of electromagnetic radiation is independent of the motion of inertial observers. It is natural to extend this circumstance to all observers, but one encounters a difficulty within the standard framework of relativity theory. However, it can be implemented within the context of the nonlocal theory of accelerated observers and plays an important role in the determination of the nonlocal kernel. 

This paper is about the relativistic theory of field measurement and the corresponding observers in spacetime. Properties of observers are discussed in section II and the consequences of nonlocality of field measurement are explored in section III. To illustrate nonlocal special relativity, the measurement of an incident electromagnetic wave by non-uniformly rotating observers is described in section IV. Finally, section V contains concluding remarks. 

\section{Observers and their characteristics}

Observers are classical macrophysical entities---either sentient or  measuring devices. Therefore, as in Newtonian physics, one can imagine the system as consisting of a collection of Newtonian point particles such that the behavior of the system can be analyzed in terms of the motion of each individual point particle. We emphasize that such a pointlike observer should still be considered a macrophysical entity subject to the laws of classical (i.e., nonquantum) physics. Therefore, in the Newtonian picture, an elementary observer is considered a point particle following a path in space. In the spacetime picture, such an observer is endowed with a local tetrad frame as well. That is, the Newtonian description can be essentially generalized by the addition of a local spatial frame of reference. Thus an elementary observer can be described via a future directed timelike world line with a suitable tetrad frame defined along the path. A laboratory would then correspond to a congruence of such world lines and their associated tetrad frames. This \emph{minimal} description of observers goes beyond the traditional one that involves only a future directed timelike world line in Lorentzian and Riemann-Cartan spacetimes~\cite{1, 2}.

The \emph{theoretical} association of a moving frame with an observer actually dates back to the early days of general relativity theory; however, this connection is now no longer a figment of imagination, but is in fact indispensable due to developments in \emph{experimental} physics. Indeed, modern measurements in spacetime generally involve the determination of tensorial or spinorial entities and this necessitates the association of a spatial frame to each elementary observer along its world line. Imagine, for example, a Paik gravity gradiometer on a space platform in orbit about the Earth~\cite{3}. The gradiometer registers inertial and gravitational forces. To separate out the effects of the translational acceleration as well as the rotation of the gradiometer from the tidal gravitational effects of the Earth, the spatial frame of the gradiometer must be controlled with sufficient accuracy~\cite{4, 5}.

We therefore define an elementary observer $\mathcal{O}$ to be given by a future directed world line $\Gamma$ and the associated orthonormal tetrad frame $\lambda^{\mu}{}_{(\alpha)}$ attached to each event characterized by the proper time $\tau$ along the world line:
 \beq \label{1}
\mathcal{O}= (\Gamma: x^{\mu}(\tau), \lambda^{\mu}{}_{(\alpha)}(\tau))\,,
\eeq
where $x^\mu=(ct, \mathbf{x})$ and $g_{\mu \nu}\lambda^{\mu}{}_{(\alpha)}\lambda^{\nu}{}_{(\beta)}=\eta_{(\alpha)(\beta)}$ signifies that the tetrad frame is orthonormal. In our convention, the signature of the spacetime metric is $+2$; moreover, we use units such that $c=1$, unless specified otherwise.
 
Thus the observer moves not just in spacetime, but in the ten dimensional \emph{state space} that is part of the \emph{frame bundle}. That is, to describe the state of an observer at each instant of proper time $\tau$, one needs an event on its future directed world line as well as its spatial frame at that event. The four coordinates of the event together with the six independent components of the frame (i.e., 16 tetrad elements subject to 10 orthonormality conditions) renders the state space a ten dimensional manifold for a general observer. 

More explicitly, the Newtonian state of an elementary observer $(\mathbf{x}(t), \mathbf{v}(t))$ is generalized here to $(t(\tau), \mathbf{x}(\tau), \mathbf{v}(\tau), \theta(\tau), \phi(\tau), \psi(\tau))$, where $(\theta, \phi, \psi)$ are the traditional Euler angles that specify the actual orientation of the spatial frame of the observer with respect to a fiducial spatial frame. The variation of the state of the observer along its trajectory can be expressed as 
 \beq \label{2}
\frac{dx^{\mu}}{d\tau}= \lambda^{\mu}{}_{(0)}\,,
\eeq
\beq \label{3}
\frac{d \lambda^{\mu}{}_{(\alpha)}}{d\tau}= \Phi_{(\alpha)}{}^{(\beta)} \lambda^{\mu}{}_{(\beta)}\,,
\eeq
where $\lambda^{\mu}{}_{(0)}(\tau)$ is the vector tangent to the world line $\Gamma$ at proper time $\tau$ and $\Phi_{(\alpha)(\beta)}$ is an antisymmetric tensor due to the orthonormality of the tetrad frame. This invariant acceleration tensor can be decomposed into spacetime scalars as $\Phi_{(\alpha)( \beta)}\mapsto (-\mathbf{g},\boldsymbol{\Omega})$ in close analogy with the decomposition of the Faraday tensor $F_{\mu\nu} \mapsto (\mathbf{E}, \mathbf{B})$ into electric $(\mathbf{E})$ and magnetic $(\mathbf{B})$ fields; that is, $F_{0i}=-E_i$ and $F_{ij}=\epsilon_{ijk}B^k$ in our convention. It follows from Eqs.~\eqref{2} and~\eqref{3} that
 the \emph{translational acceleration} of the observer is thus given by the ``electric'' part of the acceleration tensor ($\Phi_{(0)(i)}=g_{(i)}$), while the \emph{angular velocity of the rotation} of the observer's spatial frame with respect to a locally nonrotating (i.e., Fermi-Walker transported) frame is represented by the ``magnetic'' part of the acceleration tensor ($\Phi_{(i)(j)}=\epsilon_{(i)(j)(k)}\Omega ^{(k)}$)~\cite{6}. 

The spacetime invariants $\mathbf{g}(\tau)$ and $\boldsymbol{\Omega}(\tau)$ in general depend upon the instantaneous \emph{speed} of the observer as well as the orientation of its spatial frame. Under a local Lorentz transformation of the observer's tetrad frame, $\Phi_{(\alpha)( \beta)}$ transforms as a tensor. It is therefore natural to define the observer's characteristic \emph{acceleration scales} (of length and time) using the local Lorentz invariants $\mathcal{I}$ and $\mathcal{I^*}$ of the acceleration tensor~\cite{6}
\beq \label{4}
\mathcal{I}=\frac{1}{4} \Phi_{(\alpha)(\beta)} \Phi^{(\alpha)(\beta)}\,, \qquad \mathcal{I^*}=\frac{1}{4} \Phi^*_{(\alpha)(\beta)} \Phi^{(\alpha)(\beta)}\,.
\eeq
Here $\Phi^*_{(\alpha)(\beta)}$ is the dual acceleration tensor given by
\beq \label{5}
\Phi^*_{(\alpha)(\beta)}=\frac{1}{2} \epsilon_{(\alpha)(\beta)(\gamma)(\delta)} \Phi^{(\gamma)(\delta)}\,,
\eeq
where $\epsilon_{0123}=1$ in our convention. Thus we have
\beq \label{6}
\mathcal{I}=\frac{1}{2} (-g^2+\Omega^2), \qquad \mathcal{I^*} = -\mathbf{g} \cdot \boldsymbol{\Omega}\,.
\eeq
Hence, the \emph{proper} acceleration scales are usually defined via $|\mathcal{I}|^{-1/2}$ and  $|\mathcal{I^*}|^{-1/2}$~\cite{6}. We consider only positive square roots in this paper. For Earth-bound observers,  $c^{2}/|\mathbf{g}_{\oplus}|\approx 1$ light year and $c/|\boldsymbol{\Omega}_{\oplus}|\approx 28$ astronomical units. These astronomical lengths---approximately $10^{16}$ m and $5\times 10^{12}$ m, respectively---are rather large in comparison with the dimensions of a laboratory on the Earth; indeed, this circumstance is consistent with the fact that the hypothesis of locality in relativistic physics is ordinarily a very good approximation~\cite{6}.

It is interesting to note that 
\beq \label{6a}
\Phi_{(\alpha)(\gamma)} \Phi_{(\beta)}{}^{(\gamma)}=\mathcal{I}~\eta_{(\alpha)(\beta)}+4\pi \mathcal{T}_{(\alpha)(\beta)}\,, 
\eeq
\beq \label{6b}
 \Phi^*_{(\alpha)(\gamma)} \Phi_{(\beta)}{}^{(\gamma)}=\mathcal{I^*}~\eta_{(\alpha)(\beta)}\,,
\eeq
where the symmetric and traceless tensor $\mathcal{T}_{(\alpha)(\beta)}$ can be simply obtained from the electromagnetic energy-momentum tensor~\cite{7} by replacing $\mathbf{E}$ with $-\mathbf{g}$ and $\mathbf{B}$ with $\boldsymbol{\Omega}$. The general significance of such relations for the electromagnetic field have been investigated in Ref.~\cite{8}.

The notion of an elementary observer can then be extended to a reference system by considering a congruence of elementary observers that occupy a finite spacetime domain in a global inertial frame in Minkowski spacetime. That this construction is indeed possible has been demonstrated in various ways by explicit examples for simple accelerated systems~\cite{8,9,10,11}. A general method based on fiber bundles for the construction of such reference systems involving nonintegrable anholonomic observers has been discussed by Auchmann and Kurz~\cite{12}.

It is possible to provide a detailed pointwise classification of the various standard forms of the acceleration tensor $\Phi_{(\alpha)(\beta)}$ in complete analogy with electrodynamics~\cite{7, 13}. Let us denote an eigenvector and the corresponding eigenvalue of the acceleration tensor by $\Psi^{(\alpha)}$ and $\chi$, respectively. Then,
\beq \label{7}
 \Phi_{(\alpha)(\beta)}\Psi^{(\beta)}= \chi \Psi_{(\alpha)}\,.
\eeq
If $\chi \ne 0$, the associated eigenvector is null; moreover, it follows from Eq.~\eqref{7} that 
\beq \label{8}
 \chi^2= - \mathcal{I}\pm (\mathcal{I}^2+\mathcal{I^*}^2)^{1/2}\,.
\eeq
If the invariants $\mathcal{I}$ and $\mathcal{I^*}$ both vanish, then $\chi=0$ and $\Phi_{(\alpha)(\beta)}$ represents the acceleration tensor of an observer with \emph{null acceleration}~\cite{14}. If the observer's acceleration is not \emph{null}, a local Lorentz boost can always render the translational and rotational acceleration vectors \emph{parallel}. A more complete discussion along these lines would be straightforward, but beyond the scope of this paper. It is useful here to correct some minor errors that have occurred in previous publications: In Ref.~\cite{6}, Eq. (5) contains an erroneous numerical factor in the definition of $I_2$: $\frac{1}{2}$ should be replaced by $\frac{1}{4}$, and essentially the same correction is necessary in Ref.~\cite{10}. Moreover, in Ref.~\cite{14}, for the definition of $I^*$ in Eq. (3), again the numerical factor $\frac{1}{2}$ should be replaced by $\frac{1}{4}$, and in Eq. (11), $I$ should be replaced by $\frac{1}{2}I$. The main conclusions remain unchanged.

Relativistic kinematics of accelerated systems has been extensively studied for uniform rotation~\cite{7} and hyperbolic motion~\cite{14a}. The latter is the direct relativistic generalization of one-dimensional motion in Newtonian mechanics with constant acceleration. In these cases of uniformly accelerated motion, the acceleration tensor $\Phi_{(\alpha)(\beta)}$ is constant. It is possible to extend the definition of uniform acceleration to more general configurations~\cite{14b}.

It is natural to extend these general considerations to a gravitational field via Einstein's principle of equivalence. In curved spacetime, or in curvilinear coordinates in Minkowski spacetime, differentiation of the tetrad in Eq.~\eqref{3} should be replaced by covariant differentiation. It is possible to construct  \emph{nonrotating} accelerated reference systems in Minkowski spacetime, but this is in general impossible in a gravitational field, where Fermi-Walker transport can be implemented only along a single observer's world line. The nonzero spacetime curvature in general prevents the extension of the criterion for nonrotation to a congruence of observers~\cite{15}. 

If $\Phi_{(\alpha)(\beta)}(\tau)=0$ for all $\tau$, the observer is \emph{inertial}; otherwise, the observer is \emph{accelerated}. In this way, ambiguities that exist in the traditional description of observers disappear. Consider, for instance, a global inertial frame with coordinates $(t, x, y, z)$ in Minkowski spacetime and a frame of reference rotating about the $z$ axis. The rotation is turned on at $t=0$. That is, for $t \ge 0$, we have a rotating frame with coordinates $(x', y', z')$ such that $z'=z$ and
\beq \label{9}
x'=x \cos \varphi + y \sin \varphi,  \qquad  y'=-x \sin \varphi + y\cos \varphi\,.
\eeq
Here $\varphi$ is the azimuthal angle defined by
\beq \label{10}
\varphi= \int_0^t {\tilde \Omega_0}(t') dt' \,,
\eeq
so that $\varphi=0$ at $t=0$. In our convention, a tilde over the angular velocity indicates that it is expressed as a function of coordinate time rather than proper time of the observer. We are interested in the measurements of observers that are \emph{at rest} all along the $z' = z$ axis.
According to the traditional view, these observers are all \emph{inertial}; however, this viewpoint is too restrictive and could be erroneous in some circumstances. If such an observer refers its measurements to the inertial $(x, y, z)$ axes, then the observer is inertial. On the other hand, the observer may decide to refer its measurements to the rotating $(x', y', z')$ axes, in which case the observer is noninertial. The issue of whether the \emph{pointlike} observer itself rotates is immaterial and, in any case, irrelevant to the physics at hand. 

Let us now imagine that the noninertial static observers in the above example undergo accelerated motion up or down the $z'$ axis. The acceleration tensor in this case has the simple form in which the translational and rotational acceleration vectors are \emph{parallel}; that is, this situation illustrates the typical case where $\mathbf{g}\times \boldsymbol{\Omega}=0$.

Inertial observers in general follow geodesics and carry a parallel transported frame. Actual observers are all more or less accelerated. To describe measurements of accelerated observers, we need to establish a link between accelerated and inertial observers. The simplest assumption would involve \emph{local equivalence}, namely, the supposition that an accelerated observer is pointwise inertial. This \emph{hypothesis of locality} underlies the special and general theories of relativity~\cite{6, 11}. Thus Lorentz invariance can be employed at each event along an observer's world line, since the noninertial observer is assumed to be physically equivalent to an otherwise identical hypothetical inertial observer at that event. Here the inertial observer is momentarily comoving with the noninertial observer and shares the same state, so that their tetrads instantaneously coincide. That is, we imagine that along its world line, the accelerated observer together with its tetrad passes through an infinite sequence of momentarily equivalent hypothetical  inertial observers, each instantaneously comoving with the accelerated observer~\cite{6, 11}. 
 
As a concrete example, imagine the measurement of an electromagnetic field by the static observers along the $z'=z$ axis discussed above. It appears reasonable to suppose that in this case, the 3D vectors ($\mathbf{E}, \mathbf{B}$) are geometric invariants, but their measured components, namely,  their projections on the axes of the background inertial and noninertial observers are different. For instance, for the noninertial observers $E'_1 = \mathbf{E}\cdot \widehat{\mathbf{x'}}$, etc., where the rotating unit basis vectors are $\widehat{\mathbf{z'}} = \widehat{\mathbf{z}}$ and 
\beq \label{11}
\widehat{\mathbf{x'}}=\widehat{\mathbf{x}} \cos \varphi + \widehat{\mathbf{y}} \sin \varphi,  \qquad  \widehat{\mathbf{y'}}=-\widehat{\mathbf{x}} \sin \varphi + \widehat{\mathbf{y}} \cos \varphi\,.
\eeq
The corresponding components of the electric field are thus related to each other via $E'_3=E_3$ and
\beq \label{12}
E'_1=E_1 \cos \varphi + E_2 \sin \varphi,  \qquad  E'_2=-E_1 \sin \varphi + E_2 \cos \varphi\,,
\eeq
and similarly for the components of the magnetic field. These statements are essentially equivalent to the relation
\beq \label{13}
F'_{(\alpha)(\beta)}= F_{\mu\nu} \lambda^{\mu}{}_{(\alpha)} \lambda^{\nu}{}_{(\beta)}\,,
\eeq
which follows from the special theory of relativity.

On the other hand, at any given time $t$, imagine an inertial system of coordinates that has fixed spatial axes different from the background inertial system by a constant \emph{passive} rotation about the $z$ axis by $\varphi(t)$; the field components measured by such inertial observers would then coincide with those measured by the noninertial observers at time $t$. This pointwise connection between the noninertial and inertial observers may be expressed by saying that in time the rotating axes of the noninertial observers pass through an infinite sequence of passively rotated inertial systems. This example illustrates the locality hypothesis of the special relativity theory. We point out in the next section that these results must be amended, since the field measurements of noninertial observers are nonlocal.

An important consequence of the hypothesis of locality in connection with the phenomenon of spin-rotation coupling should be noted here. Consider the noninertial observers at rest along the $z'$ axis and assume that they refer their observations to axes rotating at \emph{constant} frequency ${\tilde \Omega_0}$.  Next, imagine an incident right circularly polarized plane electromagnetic wave of frequency $\omega$ and wave vector $\mathbf{k}$ with $\omega=c|\mathbf{k}|$ incident along the $z'=z$ axis. According to the inertial observers at rest along the $z$ axis, the electric and magnetic fields of the incident wave rotate with frequency $\omega$ in the positive sense about the direction of wave propagation. However, according to the static noninertial observers under consideration here, the electric and magnetic fields rotate about the direction of wave propagation with frequency $\omega-{\tilde \Omega_0}$. Thus by a mere rotation of frequency ${\tilde \Omega_0}=\omega$, the incident $\mathbf{E}$ field will be constant in time and aligned, say, with the $x'$ axis, the incident $\mathbf{B}$ field will be constant in time and aligned with the $y'$ axis and the whole electromagnetic field would thus lose its temporal dependence with respect to such noninertial observers. Indeed, for these observers the field configuration will be sinusoidal in space along the $z'$ axis, but would stand completely still. This situation, where an observer could be comoving with a radiation field, is reminiscent of the defect in the pre-relativistic Doppler formula noted by Einstein in his autobiographical notes---see page 53 of Ref.~\cite{16}. That is, an inertial observer with speed $v=c$ could be comoving with an electromagnetic wave, since in the pre-relativistic Doppler formula the observed frequency would be given by $\omega'=\omega-\mathbf{v}\cdot \mathbf{k}$, and $\omega'$ would then vanish for inertial observers pursuing the wave with speed $v=c$. This difficulty is resolved in the relativity theory of inertial observers on the  basis of Lorentz invariance, where inertial observers can approach the ultimate speed $c$, but they can never reach it. In the case of rotation, however, there is no evident upper limit to the frequency of light or the angular velocity of rotation. The two situations are similar, since they involve uniform linear and circular motions. Nevertheless, we expect that the resolution of the difficulty associated with the locality assumption will be of a totally different character; in particular, no symmetry principle is involved in noninertial motion. 
The nonlocal theory of accelerated observers in Minkowski spacetime has been so constructed as to make it impossible for \emph{any} observer to be comoving with a basic radiation field~\cite{17}. 
 
The fundamental quantum laws of microphysics have been formulated with respect to ideal inertial observers in Minkowski spacetime. It is further presumed that these observers carry ideal (i.e., perfectly sensitive and flawless) measuring devices that remain at rest in some global inertial frame of reference. We are here interested in the measurements of \emph{accelerated} observers that carry such perfectly \emph{robust} measuring devices. That is, we require further that the ideal devices carried by accelerated observers be such that during a measurement, the \emph{internal} inertial effects affecting the performance of such a device integrate to a completely negligible influence on the outcome of the measurement. Such ideal devices are called ``standard" in the theory of relativity. By definition, \emph{standard} devices are consistent with the \emph{hypothesis of locality}. Similarly, in the discussion of acceleration-induced nonlocality, it is important to assume that all measuring devices are standard as well~\cite{18}. In fact, as will become clear in the next section, this assumption is necessary for a proper formulation of the acceleration-dependent nonlocal kernel, which characterizes the observer's memory of accelerating through the Minkowski vacuum.

\section{Bohr-Rosenfeld Principle: Nonlocality of Field Measurement}

Huygens' principle suggests that in general wave properties cannot be measured at one event. Various thought experiments involving measurement of electromagnetic wave phenomena by accelerated observers indicate that for wave phenomena the hypothesis of locality is valid only in the ray limit, where $\lambda/\mathcal{L}$ is negligible. Here $\lambda$ is the characteristic wavelength of the phenomenon under observation and $\mathcal{L}$ is the relevant acceleration length. We therefore expect that in this case deviations from locality would be proportional to $\lambda/\mathcal{L}$~\cite{18}.

Bohr and Rosenfeld~\cite{19, 20} have pointed out that the measurement of the electromagnetic field by ideal inertial observers involves an averaging process over a past spacetime domain. A direct extension of the Bohr-Rosenfeld treatment to ideal accelerated observers appears to be a rather daunting task; instead, we approach this basic problem indirectly as follows. Let us first note that for field measurements by actual (i.e., accelerated) observers, their finite acceleration scales must be taken into account as well. When extended to an \emph{elementary noninertial observer}, the Bohr-Rosenfeld \emph{principle}~\cite{21} simply implies that the memory of the field along the past world line of the observer must be taken into account. That is, the averaging process reduces to an integration over its past world line, as the elementary observer has no spatial extension by assumption and is thus represented only by its world line and the associated frame. Thus in searching for a physical link between a noninertial observer and the class of hypothetical momentarily comoving inertial observers along its past world line, we must go beyond the pointwise condition~\eqref{13} and consider a nonlocal relationship involving the past world line of the observer. To simplify matters, we demand that the nonlocal connection be \emph{linear}; indeed, this requirement is consistent with the Bohr-Rosenfeld \emph{averaging} procedure. The most general linear relation consistent with causality is of the form~\cite{NAO}
\begin{eqnarray}\label{14}
\mathcal{F}_{(\alpha)(\beta)}(\tau)=F'_{(\alpha)(\beta)}(\tau)+u(\tau-\tau_{0})\int_{\tau_{0}}^{\tau}K_{(\alpha)(\beta)}{}^{(\gamma)(\delta)}(\tau,\tau')
F'_{(\gamma)(\delta)}(\tau')d\tau'\,,
\end{eqnarray}
where $\mathcal{F}_{(\alpha)(\beta)}$ is the field measured by the accelerated observer at time $\tau$ and the kernel is expected to be proportional to the acceleration tensor $\Phi_{(\alpha)(\beta)}$. Here, $u(t)$ is the unit step function such that $u(t) = 0$ for $t<0$ and $u(t) = 1$ for $t>0$ and $\tau_{0}$ is the instant of proper time at which the observer's acceleration is initially turned on. For a radiation field $F_{\mu \nu}$ in the eikonal approximation, we expect that the nonlocal term in Eq.~\eqref{14} would be proportional to $\lambda/\mathcal{L}$, so that locality is recovered in the eikonal limit. 

Ansatz~\eqref{14} introduces a new basic element into relativity theory, namely, the acceleration-dependent nonlocal kernel that weighs the significance of past events for the present time. The general physical content of Eq.~\eqref{14} for a basic field may be expressed as follows: For classical point particles and rays of radiation, the measurements of actual (accelerated) observers are pointlike and the locality hypothesis is valid; however, for field measurements, observers have \emph{memory}~\cite{22} and are in general nonlocal. This nonlocality has a special character that must be clarified here; that is, at time $\tau$ the accelerated observer measures a local field $\mathcal{F}_{\mu \nu}(x)$ by projecting this field on its local tetrad frame; however, the local field is a solution of nonlocal integro-differential field equations.  In the case of ideal inertial observers, their acceleration scales are all infinite and though their field measurements would involve averaging in accordance with the considerations of Bohr and Rosenfeld~\cite{19, 20}, such averaging is innocuous in \emph{classical field theory} in the absence of finite acceleration scales  and can be essentially taken to the pointlike limit. 

It is possible to think of Eq.~\eqref{14} as an expansion of the measured field in powers of $F_{\mu \nu}$, where the terms beyond the linear order have been simply neglected. The provisional character of our ansatz should thus be emphasized. Perhaps future observational data will provide the necessary motivation to go beyond the present linear theory. 

The approach to relativity theory in Minkowski spacetime outlined here leads to nonlocal special relativity~\cite{17}, where we assume measuring devices are all standard in order that the explicit effects of acceleration would only appear in the nonlocal kernels. These originate from the circumstance that the determination of physical fields is \emph{not} instantaneous and requires measurements along the past world line of the observer. Thus in a measured field, the acceleration of the world line is explicitly represented by a universal kernel that acts as the weight function for the memory of past acceleration. 

Let us digress here and mention briefly the status of observers in nonlocal gravity. The principle of equivalence implies a profound connection between inertia and gravitation. In analogy with acceleration-induced nonlocality, it is possible to develop a nonlocal generalization of Einstein's theory of gravitation via the teleparallel equivalent of general relativity~\cite{21a, 21b, 21c, 21d}. In this \emph{tetrad} theory of nonlocal gravity, gravitation is described by a local field that satisfies nonlocal integro-differential field equations such that nonlocality simulates dark matter. Moreover, as before, observers follow future directed timelike world lines and carry orthonormal tetrad frames. Though all measurements involve local fields, observers in a gravitational field are intrinsically nonlocal as a natural result of the nonlocality of the gravitational field equations. 

The existence of the acceleration-induced nonlocal kernel for field measurements in Minkowski spacetime must be regarded as a purely \emph{vacuum} effect of accelerated motion, since all measuring devices have been assumed to be \emph{standard}. The main task of nonlocal special relativity is the determination of nonlocal kernel for various fields. The current status of this general problem has been briefly reviewed in Ref.~\cite{21}; therefore, we confine our considerations here to the electromagnetic kernel in Eq.~\eqref{14}. The first step involves the assumption that~\cite{8, 22}
\begin{eqnarray}\label{15}
K_{(\alpha)(\beta)}{}^{(\gamma)(\delta)}(\tau,\tau')=k_{(\alpha)(\beta)}{}^{(\gamma)(\delta)}(\tau')\,,
\end{eqnarray}
which corresponds to \emph{kinetic} memory in the terminology of Ref.~\cite{22}. An important property of kernels of this type is that after the observer's acceleration has been turned off and the motion is uniform, the field simply carries a \emph{constant} memory of past acceleration. For instance, in Eq.~\eqref{14}, the nonlocal part would become a constant once the acceleration has been turned off. Such constant fields, which contain the influence of past accelerations, can be canceled in a measuring device when it is reset.

For a pure radiation field, the kernel can be determined from the physical requirement that no observer can be comoving with the field~\cite{17, 21}. However, in electrostatics and magnetostatics, electrodynamics deals with constant fields as well. Thus in Eq.~\eqref{14}, the next step in the determination of the kernel would involve the consideration of all possible local combinations of the acceleration tensor, the Minkowski metric tensor and the Levi-Civita tensor that could generate kernels of the form given in Eq.~\eqref{15}. We assume here for the sake of simplicity that the kernel is \emph{linearly} dependent upon the acceleration of the observer and this reduces the possible forms of the kernel to  
\begin{equation}\label{16}
\kappa_{(\alpha)(\beta)}{}^{(\gamma)(\delta)}=-2\Phi_{\lbrack(\alpha)}{}^{\lbrack(\gamma)}~\delta_{(\beta)\rbrack}{}^{(\delta)\rbrack}
\end{equation}
and its dual given by~\cite{21}
\begin{equation}\label{17}
\kappa^*_{(\alpha)(\beta)}{}^{(\gamma)(\delta)}=-2\Phi^*_{\lbrack(\alpha)}{}^{\lbrack(\gamma)}~\delta_{(\beta)\rbrack}{}^{(\delta)\rbrack}\,.
\end{equation}
Finally, to simplify matters further, we assume that the kernel is given by a linear combination of Eqs.~\eqref{16} and~\eqref{17} with \emph{constant} dimensionless coefficients $p$ and $q$, respectively. Various properties of such a kernel have been discussed in Refs.~\cite{17, 21}. For $p=1$ and $q=0$, kernel~\eqref{16} would simply correspond to that of a pure radiation field; however, as discussed in detail in Refs.~\cite{17, 21}, we expect that $p\ge0$ and $0<|q|\ll1$ for the kernel in Eq.~\eqref{14}. The presence of $q$ indicates possible violations of invariance under time reversal and parity in an accelerated system.

The Bohr-Rosenfeld principle and the physical arguments that demonstrate the limited validity of the hypothesis of locality provide the necessary motivation to amend the standard theory. However, the generalization of the locality assumption that is contained in Eq.~\eqref{14}, together with similar relations for other basic fields, is the crucial step that leads to acceleration-induced nonlocality. Is this step verifiable? Of course, the gist of this  step is the existence of the nonlocal kernel and one can look for the observable consequences of the existence of such a kernel. For instance, a prediction of the nonlocal theory that is consistent with current observational data should be mentioned here: Nonlocal special relativity forbids the existence of scalar or pseudoscalar radiation fields in nature. Moreover, regarding nonlocal electrodynamics of accelerated systems under consideration in this paper, there is a lamentable lack of reliable observational data; therefore, one can contemplate indirect confirmation of this theory via the ``accelerated" motion of electrons in the correspondence limit of quantum mechanics. That is, electrons undergoing ``circular" motion in the limit of large quantum numbers are expected to behave much like accelerated observers. A detailed investigation reveals that the implications of the nonlocal theory are in better qualitative agreement with quantum theory than the standard local theory~\cite{23}. At present, one can only hope that future experiments will directly verify the main tenets of nonlocal special relativity. 

Some of the observational consequences of Eqs.~\eqref{14}--\eqref{17} have been explored in previous publications; in particular, the implications of this theory of nonlocal electrodynamics for the phenomenon of spin-rotation coupling have been studied for \emph{uniformly rotating} systems in Refs.~\cite{17, 21}. We therefore consider the case of \emph{non-uniform rotation} in the next section. 

\section{Non-Uniformly Rotating Observers}

In the rotating $(t, x', y', z')$ coordinate system of section II, imagine the class of fundamental observers at rest in space. In a cylindrical system of spatial coordinates $(r, \vartheta, z)$, each such fundamental observer can be characterized by fixed values of these rotating coordinates, and hence by $(t, r\cos \Theta, r\sin \Theta, z)$ with respect to the inertial system of coordinates, where $\Theta=\vartheta+\varphi(t)$. The proper time along the path of such an observer is given by
\begin{equation}\label{1C}
\tau=\int_0^t[1- \tilde{v}^2(t')]^{1/2}~dt'\,,
\end{equation}
where $\tilde{v}(t)=r{\tilde \Omega_0}(t)$ is the azimuthal speed of motion of the observer and we have assumed here that $\tau=0$ at $t=0$. Let $\Omega_0(\tau):={\tilde \Omega_0}(t)$, $v(\tau):=\tilde{v}(t)$ and $\gamma=dt/d\tau$ be the corresponding Lorentz factor; then, the natural tetrad frame of such a non-uniformly rotating observer is given with respect to the inertial $(t, x, y, z)$ coordinate system by~\cite{11, 24}
\begin{align}\label{2C} \lambda^\mu_{\;\;(0)} &=\gamma (1,-v\sin\Theta ,v\cos \Theta ,0),\\
\label{3C}\lambda^\mu_{\;\;(1)}&=(0,\cos \Theta ,\sin \Theta ,0),\\
\label{4C} \lambda^\mu _{\;\;(2)}&=\gamma (v,-\sin \Theta ,\cos \Theta ,0),\\
\label{5C}\lambda^\mu _{\;\;(3)}&=(0,0,0,1).\end{align}
It follows from Eq.~\eqref{3} that the translational and rotational acceleration vectors of a typical rotating observer are given by
\begin{equation}\label{6C}
\mathbf{g}=(-v\gamma^2\Omega_0, \gamma^2\frac{dv}{d\tau},0)\,,
\end{equation}
\begin{equation}\label{7C}
\mathbf{\Omega} =(0, 0, \gamma^2\Omega_0)\,,
\end{equation}
where the components of these vectors are expressed here with respect to local spatial axes of the observer's frame $\lambda^\mu{}_{(i)}, i=1, 2, 3$, that indicate the radial, tangential and $z$ directions, respectively. As expected, the translational acceleration vector has only centripetal and tangential components in this case, so that $\mathcal{I^*}=0$. If the angular velocity of rotation is so chosen that $\mathcal{I}=0$ as well, then we have the case of null accelerated observers discussed in Ref.~\cite{11}.

It is interesting to investigate the influence of tangential acceleration $\gamma^2 dv/d\tau$ on the phenomenon of spin-rotation coupling for an incident electromagnetic wave. Let us therefore consider a plane monochromatic wave of frequency $\omega$ that is incident along the $z$ axis. We represent this wave by a column 6-vector $F$ whose components are $\mathbf{E}$ and $\mathbf{B}$, respectively; that is, 
\begin{equation}\label{8C}
 F_\pm (t,\mathbf{x})=f_{\pm} \begin{bmatrix} \mathbf{e}_\pm\\ \mathbf{b}_\pm\end{bmatrix} e^{-i\omega (t-z)}\,, 
\end{equation}
where the upper (lower) sign represents positive (negative) helicity radiation and $f_{+}$ and $f_{-}$ are the corresponding constant amplitudes. The unit circular polarization vectors 
\begin{equation}\label{9C}
\mathbf{e_\pm}=(\widehat{\mathbf{x}} \pm i\widehat{\mathbf{y}} )/ \sqrt2\,, \qquad \mathbf{b}_\pm =\mp i\mathbf{e}_\pm\,,
\end{equation}
are such that $\mathbf{e}_{\pm} \cdot \mathbf{e}^*_{\pm}=1$. It is convenient to employ complex fields and adopt the standard convention that only their real parts have physical significance, since the basic nonlocal ansatz~\eqref{14} is linear.

It follows from Eq.~\eqref{13} that the field measured by a typical momentarily comoving inertial observer along the world line for $\tau \ge 0$ is given by
\begin{equation}\label{10C}
 F'_\pm (\tau)=\gamma f_{\pm} \begin{bmatrix} \mathbf{e'}_\pm\\ \mathbf{b'}_\pm\end{bmatrix} e^{i\eta}\,, 
\end{equation}
where 
\begin{equation}\label{11C}
\eta=-\omega \int_0^{\tau}\gamma(\tau')~d\tau' +\omega z \pm \Theta(\tau)\,. 
\end{equation}
Here $\mathbf{b'}_\pm =\mp i\mathbf{e'}_\pm$ and 
\begin{equation}\label{12C} 
\mathbf{e'}_\pm =\frac{1}{\sqrt2} \begin{bmatrix} 1\\ \pm i\gamma^{-1}\\ \pm iv \end{bmatrix}\,.\end{equation}
These components refer to the local spatial frame of the observer, namely, $\lambda^{\mu}{}_{(i)}(\tau), i=1, 2, 3$, and form unit circular polarization vectors such that $\mathbf{e'}_{\pm} \cdot \mathbf{e'}^*_{\pm}=1$; moreover, they reduce to those in Eq.~\eqref{9C} for $v=0$. This means that the comoving inertial observers perceive circularly polarized radiation of effective ``frequency" $\omega'$, 
\begin{equation}\label{13C} 
\omega'(\tau)=\gamma[\omega \mp \Omega_0(\tau)]\,,
\end{equation}
in terms of which the phase of the wave can be expressed as
\begin{equation}\label{14C}
\eta=- \int_0^{\tau}\omega'(\tau')~d\tau' +\omega z \pm \vartheta\,. 
\end{equation}

The incident electromagnetic field measured by the accelerated observer is given by Eq.~\eqref{14}, which can be thought of as a matrix relation with a kernel that is a $6\times 6$ matrix. The simple provisional kernel discussed in the previous section can be expressed in matrix notation as
\begin{equation}\label{15C}
 k=p\;\kappa +q\;\kappa^*\,,
 \end{equation}
where
\begin{equation}\label{16C}
 \kappa =\begin{bmatrix} \kappa_1 & -\kappa_2\\ \kappa_2 &\kappa_1 \end{bmatrix} ,\quad \kappa^* =\begin{bmatrix} -\kappa_2 & -\kappa _1\\ \kappa_1 & -\kappa_2 \end{bmatrix} .\end{equation}
It follows from Eqs.~\eqref{16} and~\eqref{17} that 
\begin{equation}\label{17C}
\kappa_1 =\mathbf{\Omega} \cdot\mathbf{I}\,, \qquad \kappa_2 =\mathbf{g}\cdot \mathbf{I}\,, 
\end{equation}
where $I_i$, with components $(I_i)_{jk}=-\epsilon _{ijk}$, is a $3\times3$ matrix that is proportional to the operator of infinitesimal rotations about the $x^i$ axis.

The incident electromagnetic field as measured by the accelerated observer is then 
\begin{equation}\label{18C} 
\mathcal{F}_\pm (\tau)=F'_\pm (\tau ) -(p \pm iq)f_\pm \int_0^\tau \gamma(\tau') \Sigma(\tau')e^{i\eta(\tau')}~d\tau'\,,
\end{equation}
where $\Sigma$ is given by
\begin{equation}\label{19C}
\Sigma (\tau)=\pm i\gamma \Omega_0  \begin{bmatrix} \mathbf{e'}_\pm\\ \mathbf{b'}_\pm\end{bmatrix} + \gamma^2 \frac{dv}{d\tau}\begin{bmatrix} \mathbf{a'}_\pm\\ \mp i \mathbf{a'}_\pm\end{bmatrix}\,. 
\end{equation}
Here $\mathbf{a'}_\pm$,
\begin{equation}\label{20C} 
\mathbf{a'}_\pm =\frac{1}{\sqrt2} \begin{bmatrix}  v \\ 0 \\ \pm i \end{bmatrix}\,,
\end{equation}
are the polarization vectors associated with tangential acceleration in this case. Indeed, they depend on the helicity of the incident radiation, but there is no additional time-dependent phase factor in Eq.~\eqref{19C} and consequently no direct coupling between helicity and translational acceleration that would be similar to the spin-rotation coupling. Thus the situation regarding the relation between photon helicity and translational acceleration of the observer in this nonlocal case is analogous to the local theory~\cite{24a}. In any case, it is interesting to see---as is evident from Eq.~\eqref{19C}---that nonlocality explicitly brings out the influence of the tangential acceleration of the observer in addition to the expected contribution of the spin-rotation coupling. For constant angular velocity, $\mathcal{F}_\pm (\tau)$ reduces to the pure spin-rotation result given in Ref.~\cite{17}.

Let us now return to the uniformly rotating observers \emph{at rest} along the $z'$ axis and determine what the nonlocal theory predicts for their measured field. For these observers $v=0$, $\gamma=1$, $\tau=t$ and $\Omega_0={\tilde \Omega_0}$. Then, $\mathcal{F}_\pm (t)$ is given by $W_{\pm} (t)$ up to constant coefficients; that is, 
\begin{equation}\label{21C}
\mathcal{F}_\pm (t)=W_\pm (t)f_\pm e^{i(\omega z\pm\vartheta)} \begin{bmatrix} \mathbf{e'}_\pm\\ \mathbf{b'}_\pm\end{bmatrix}\,,
\end{equation}
where 
\begin{equation}\label{22C}
W_\pm (t)=\left [1+\frac{(\pm p+iq)\Omega_0}{\omega \mp \Omega_0} \right ]e^{-i\omega' t}-\frac{(\pm p+iq)\Omega_0}{\omega \mp \Omega_0}\,.
\end{equation}
Thus the field in general contains a harmonic component of frequency $\omega'=\omega \mp\Omega_0$ together with a constant component that is due to the boundary condition at $t=0$. The measured frequency of the circularly polarized electromagnetic wave is indeed generally the same as the perceived angular velocity of rotation of the electric and magnetic fields about the direction of wave propagation, except in the case of resonance ($\omega=\Omega_0$). The helicity-rotation coupling that is manifest in $\omega'$ has firm observational support; for instance, it accounts for the \textit{phase wrap-up} in the GPS system~\cite{25, 26}.
Moreover, in the case of resonance involving an incident \emph{positive} helicity radiation with frequency $\omega=\Omega_0$, the measured field is linearly dependent upon time with 
\begin{equation}\label{23C} 
W_{+}(t)=1-i(p+iq) \Omega _{0}t\,;
\end{equation}
in particular, the uniformly rotating observers will never be comoving with the wave, as expected. 

\section{Discussion}

In the design of electric machines, the electromagnetic field experienced by a moving part is traditionally estimated by assuming that it is instantaneously motionless~\cite{27, 28}. Thus in customary engineering applications, the instantaneous velocity, acceleration, etc., at any given time $t$ are all neglected, while in the locality hypothesis of the standard theory of relativity only the instantaneous velocity is taken into account. To obtain observational data that would pertain to an examination of the foundations of electrodynamics of accelerated systems, one must have access to experiments of rather high sensitivity. Unfortunately, however, further progress is currently hampered by a serious lack of reliable observational data---see~\cite{8, 17, 28, 29, 30} and the references cited therein.

\begin{acknowledgments}
I am grateful to C. Chicone, F. W. Hehl and S. N. Lyle for valuable discussions.

\end{acknowledgments}

\appendix

\end{document}